\documentclass[12pt]{article}
\usepackage[T1]{fontenc}
\usepackage{epsfig}

\newcommand{\chup}{$\chi U\phi$}
\begin{document}
                                                                              
%\vspace{1cm}
\begin{center}
  {\LARGE\bf Study of compact abelian lattice gauge theories} \footnote{To be
    published in the {\em\v{C}eskoslovensk\'y \v{c}asopis pro fyziku,}
    Prague (in Czech).}\\
  \vspace{15mm}
  Ji\v{r}{\'{\i}} Jers\'ak,\\
  Institute of Theoretical Physics E, RWTH Aachen, D-52056 Aachen, Germany
\end{center}
\vspace{5mm}

%\hfill DRAFT 06.10.00 \\

\begin{abstract}
  This is a review, intended for lattice nonspecialists, of the studies of the
  compact abelian gauge theories on the lattice performed by the Aachen
  lattice field theory group. We discuss in particular the pure compact QED
  and a U(1) lattice gauge theory with charged scalar and fermion fields in
  four and three dimensions. Our data indicate that these lattice theories
  might define some continuum field theories, or at least effective field
  theories with remarkable nonperturbative properties like confinement and
  dynamical mass generation.
\end{abstract}

%%%%%%%%%%%%%%%%%%%%%%%%%%%%%%%%%%%%%%%%%%%%%%%%%%%%%%%%%%%%%%%%%%%%%%%%
\section{Why to study abelian lattice gauge theories?}
%%%%%%%%%%%%%%%%%%%%%%%%%%%%%%%%%%%%%%%%%%%%%%%%%%%%%%%%%%%%%%%%%%%%%%%%

As it is well known, the formulation of quantum chromodynamics (QCD) on the
lattice substantially contributed to the theoretical understanding of
nonperturbative properties of strong interactions on the distance of the order
of one Fermi, like confinement of quarks or spontaneous breaking of chiral
symmetry, and opened way for calculation of many properties of hadrons and of
the quark-gluon plasma by means of numerical simulation. This success of
lattice QCD is due, among other things, to the asymptotic freedom of QCD,
which provides a unique prescription how to reach the limit of continuum
spacetime.  This leads naturally to the question what results have been
obtained by means of the lattice regularization and numerical simulation in
the study of nonperturbative properties of field theories that are not
asymptotically free.

In this report, I concentrate on the discussion of results obtained by our
research group in Aachen for abelian gauge theories. The most important
example of such theories is quantum electrodynamics (QED), describing
interaction of a U(1) gauge field with charged fermions. But relevant is also
electrodynamics with charged scalar field and with both such fields
simultaneously.

Though QED is phenomenologically the most successful field theory, until now
it is not clear, what are its properties at very short distances. As QED is
not asymptotically free, the running fine structure constant increases with
energy scale, and according to the perturbative results diverges in the Landau
pole. Therefore, QED is not defined by a perturbation series at energies
comparable and larger then the position of this pole. In the case of U(1)
gauge theories with other charged fields the situation is similar to QED.

What really happens around the Landau pole is a nonperturbative question. It
might seem to be irrelevant phenomenologically, since in QED the Landau pole
lies far beyond the Planck scale.  However, it is a very interesting question
from the fundamental point of view and for the search of new gauge theories
applicable to physics beyond the standard model. These reasons also motivated
our choice of the so-called compact formulation of the abelian gauge theory on
the lattice, as here the nonperturbative phenomena are particularly rich and
analogous to nonabelian theories.

%%%%%%%%%%%%%%%%%%%%%%%%%%%%%%%%%%%%%%%%%%%%%%%%%%%%%%%%%%%%%%%%%%%%%%%%
\section{Gauge theories on the lattice.}
%%%%%%%%%%%%%%%%%%%%%%%%%%%%%%%%%%%%%%%%%%%%%%%%%%%%%%%%%%%%%%%%%%%%%%%%

Lattice regularization of field theories usually uses the euclidean spacetime,
which is discretized by a hypercubic lattice with the lattice constant $a$,
corresponding to the energy-momentum cutoff $\pi/a$.

Derivatives of fields are replaced by differences of their values at the
neighbour points $(x, x+ae_\mu)$ divided by $a$. In a Lagrangian
there arise products of fields at neighbour points of the type
\begin{equation}
       \frac{1}{a} \overline{\chi}(x) \gamma^\mu U_{x,\mu} \chi(x+ae_\mu).
  \label{nn}
\end{equation}
For charged fields the gauge invariance requires an insertion of the quantity
$U_{x,\mu}$, which under the U(1) gauge transformation
$exp(\Theta(x))$ transforms as
\begin{equation}
       U_{x,\mu} \rightarrow e^{-\Theta(x)} U_{x,\mu} e^{\Theta(x+ae_\mu)},
  \label{gaugetrafo}
\end{equation}
i.e. like the parallel transporter
\begin{equation}
  exp{\left( ig_0\int_{x}^{x+ae_\mu}dy^\mu A_\mu \right)} 
\end{equation}
in continuum. Here $A_\mu$ is the vector potential and  $g_0$ is the bare
charge of the 
fermion field $\chi$. For infinitesimal $a$ one can imagine $U_{x,\mu}$
as being
\begin{equation}
                    U_{x,\mu} \cong \exp{\left( ig_0aA_\mu(x) \right)},
  \label{U}
\end{equation}
which gives the term proportional to $A_\mu$ in the covariant derivative.

Quantities $U_{x,\mu}$, called link variables, 
are on the lattice the most suitable variables for the description of the
gauge field, as they allow an easy construction of gauge invariant expressions
like (\ref{nn}) when $a$ is not
infinitesimal. In difference to  $A_\mu(x) \in (-\infty, \infty)$
the values of $U_{x,\mu}$ are from the compact manifold of the U(1)
group. This substantially simplifies the definition of the path integrals with
various functions of fields $F$,
\begin{equation}
  \int_{U(1)} D U_{x,\mu} \; \int D(other \; fields) \exp\left(
    -\frac{1}{\hbar}S \right) \; F(U_{x,\mu}, other \; fields),
  \label{PI}
\end{equation}
since it is not necessary to fix the gauge. 

Formulation of abelian gauge theories on the lattice which uses for 
description of the gauge field the link variables
$U_{x,\mu}$ is called compact. It is also possible to stay on the lattice with
the variables $A_\mu(x)$, which results in the noncompact formulation. The
difference is physically substantial at intermediate and strong coupling.

The path integral (\ref{PI}) has a form similar to a partition function of
many systems in statistical mechanics, for example of spin models on the
lattice. This similarity reflects the fundamental relationship between the
quantum field theory and statistical physics, which makes possible the use of
methods of statistical physics for a solution of nonperturbative problems in
the field theory.  This in particular includes the use of the renormalization
group in the Wilson formulation. Formally the temperature is replaced in
(\ref{PI}) by the Planck constant $\hbar$ (everywhere else we use units in
which $\hbar=c=1$), which physically means substituting thermal fluctuations
by quantum ones.

Action $S$ in the path integral (\ref{PI}) is chosen mainly on the basis of
the requirements of gauge invariance and of the usual form in the limit $a
\rightarrow 0$,
\begin{equation}
  S \rightarrow \frac{1}{4}\int d^4x F^{\mu\nu}F_{\mu\nu}.
  \label{scont}
\end{equation}
These requirements do not determine $S$ on the lattice uniquely; it is usually
expected, however, that various choices of $S$ give the same continuum limit
of the path integrals (\ref{PI}). In the language of the renormalization group
this means the same universality class. However, sometimes it happens that
different choices of $S$ or different values of parameters in $S$ lead to
different continuum limits.

Usually the choice of $S$ for gauge field is based on the simplest gauge
invariant product of link variables along the shortest closed paths formed by
the rands  $\partial p$ of elementary squares (plaquets) on the lattice,
\begin{equation}
  U_p = \prod_{\partial p} U_{x,\mu} = e^{i\Theta_p} \; \in \; U(1).
  \label{Up}
\end{equation}
Real part of this product fulfills all the formal requirements on $S$, and
therefore the choice of $S$ is, as suggested by K. Wilson,
\begin{equation}
  S = \beta \sum_p (1-\cos\Theta_p) .
  \label{S}
\end{equation}
Using (\ref{U}) it is easy to verify that
\begin{equation}
  \Theta_p \cong a^2 g_0 F_{\mu\nu},
  \label{thetap}
\end{equation}
\begin{equation}
  S \cong \beta \sum_p \left(\frac{1}{2} a^4 g_0^2 F_{\mu\nu}^2
   + a^4 O(a^4 g_0^4 F_{\mu\nu}^4) \right).
  \label{Sexpanded}
\end{equation}
Requirement (\ref{scont}) is fulfilled by the choice
\begin{equation}
  \beta = \frac{1}{g_0^2},
  \label{beta}
\end{equation}
so that $\beta$ in (\ref{S}) has the physical meaning of the inverse square of
the bare coupling constant. Perturbative expansion holds for large $\beta$.

The remarkable property of the described lattice gauge theory is the
selfinteraction of the gauge field, illustrated by the second term in the
expansion (\ref{Sexpanded}). It is a consequence of the compact formulation,
which becomes relevant for $g_0 = O(1)$ and, as we shall see, leads to
interesting and mostly still open field theoretical questions.

Formulation of nonabelian gauge theories is made in an analogous way, the
values of the link variables taken from the compact manifolds of the
corresponding gauge groups. The properties of these theories are rather simple
and well understood in the whole interval $0 < g_0 < \infty$, which is the
basis of all the success of the lattice QCD.

Formulation of the scalar field theories on the lattice is straightforward and
does not lead to substantially new insights. As for the fermions, in this
overview we use the so-called staggered fermions, fully including their
dynamics (``unquenched'' fermions). We can only mention that conceptual
problems with lattice fermions burdened the lattice field theory for a long
time. In the last years, a substantial progress has been achieved, however.
In the case of QCD the main remaining problem now is the high cost of
realistic calculations for quark fields.

%%%%%%%%%%%%%%%%%%%%%%%%%%%%%%%%%%%%%%%%%%%%%%%%%%%%%%%%%%%%%%%%%%%%%%%%%%
\section{Renormalization of field theories on the lattice}
%%%%%%%%%%%%%%%%%%%%%%%%%%%%%%%%%%%%%%%%%%%%%%%%%%%%%%%%%%%%%%%%%%%%%%%%%%

The primary issue of every lattice field theory is the procedure how to
construct the limit $a \rightarrow 0$ for ratios of path integrals of the type
(\ref{PI}), which are related to various physical observables. This
corresponds to the removal of the cutoff and thus the return to the continuum
spacetime. The procedure is a variant of the renormalization group methods. It
has been formulated by Wilson and is consequently nonperturbative. The most
important steps are the following:

Some given theory on the lattice contains parameters $K_l$.  An example is
$\beta$ in the action (\ref{S}). Multiplying by suitable powers of the lattice
constant $a$ it is possible to define all $K_l$ dimensionless.  For example, a
bare mass $m_0$ of some field enters the theory only in the form $am_0$. Then
the only dimensionful parameter is $a$. Let us imagine that for various values
of the parameters $K_l$ it is possible to calculate two- and other n-point
functions, which allow to determine masses and other observables.  For
dimensional reasons these observables must be proportional to some powers of
$a$. For example masses have the form
\begin{equation}
       m_i = \frac{1}{a \; \xi_i},
  \label{mass}
\end{equation}
where $\xi_i$ are dimensionless numbers depending on $K_l$. In statistical
mechanics these $\xi_i$ are called correlation lengths, as they determine the
asymptotic exponential decay of two-point functions with the distance $n$
between both points expressed in units of $a$,
\begin{equation}
       \langle \Phi_i(x) \Phi_i(x+nae_\mu) \rangle \; \propto \; e^{-n/\xi_i}.
  \label{corel}
\end{equation}
By inserting into this relation $\xi_i$ obtained from eq.  (\ref{mass}), we
obtain the typical asymptotic behaviour of the field theoretic propagator of a
particle of the mass $m_i$.

It is clear from eq. (\ref{mass}) that if the mass $m_i$ should have a
sensible finite value in the limit $a \rightarrow 0$, it is necessary to
assure that
\begin{equation}
            \xi_i \rightarrow \infty.
  \label{xi}
\end{equation}
This means that the value of $a$ decreases when the parameters $K_l$ are tuned
to some critical point. Thus the fundamental requirement for the construction
of a continuum limit is the existence of a continuous phase transition, i.e.
of a transition of second or higher order, for some critical values $K_l =
K_l^c$.  Phase transitions of first order do not possess a divergence of the
type (\ref{xi}).

Next important condition for renormalizability is the validity of a
scaling behaviour when the parameters approach the critical point $ K_l^c$,
\begin{equation}
               K_l \rightarrow K_l^c.
  \label{KKC}
\end{equation}
Correlation lengths in statistical physics, and thus masses in
lattice field theories, usually have the scaling behaviour of the type
\begin{equation}
 \frac{1}{\xi_i} \; = \; am_i \simeq c_i |K - K^c|^\nu, \;\;\; \nu > 0,
  \label{scalpot}
\end{equation}
or if $ K^c = \infty$,
\begin{equation}
 \frac{1}{\xi_i} \; = \; am_i \simeq c_i e^{-bK}, \;\;\; b > 0.
  \label{scalexp}
\end{equation}
Here we assume for simplicity that it is sufficient to tune one parameter
$K$. If one of the sets of relations (\ref{scalpot}),
(\ref{scalexp}) holds, it is possible to perform the limit (\ref{KKC}) for
masses with the result
\begin{equation}
  \frac{m_i}{m_1} = \frac{c_i}{c_1}.
  \label{ratio}
\end{equation}
In this way the ratios between the masses are determinded. Assuming some value
in GeV for some mass $m_1$ (taken for example from the experiment), we obtain
a prediction for the other masses in GeV in the continuum limit. In addition,
we gain the information about the size of the lattice constant $a$ in physical
units (GeV)$^{-1}$ in the vicinity of the critical point. In the case
(\ref{scalpot}) it is
\begin{equation}
  a \simeq \frac{1}{m_1} c_1 |K - K^c|^\nu.
  \label{aGeV}
\end{equation}
This is a very important information, as practical calculations, both
numerical and analytical, cannot be performed directly at the critical point,
but only in its vicinity.

Other physical quantities, like coupling constants and condensates, are
treated in a similar way.

In the case of lattice QCD the situation is, due to the asymptotic freedom,
clear and quite simple: the critical point is at $\beta^c = \infty$, and the
scaling behaviour of the type (\ref{scalexp}) holds, with a known value of the
coefficient $b$. With these certitudes it is possible in the lattice QCD to
extrapolate various quantities into the limit $a \rightarrow 0$ quite reliably
even if the used sizes of $a$ are not very small.

%%%%%%%%%%%%%%%%%%%%%%%%%%%%%%%%%%%%%%%%%%%%%%%%%%%%%%%%%%%%%%%%%%%%%%%%
\section{Review of the situation in abelian theories}
%%%%%%%%%%%%%%%%%%%%%%%%%%%%%%%%%%%%%%%%%%%%%%%%%%%%%%%%%%%%%%%%%%%%%%%%

In the case of the abelian lattice gauge theories the situation is much more
complicated. For large values of $\beta$, which according to the relation
(\ref{beta}) correspond to small values of the bare charge $g_0$, these
theories (in four-dimensional spacetime) have properties known from the
perturbative expansion in QED. For example the gauge (photon) field is
massless, static charges act on each other by the Coulomb force, and also
higher order results hold. This is satisfactory, but at the same time it means
that here the problem of the Landau pole remains the same as in continuum.

For small values of $\beta$, the properties of these theories are quite
different from those at large $\beta$. Using standard methods of statistical
physics, in particular the expansion in powers of $\beta$, usually called the
strong coupling or high temperature expansion, a lot of surprising results can
be derived. Gauge field is massive and forms states called gauge balls,
similar to the glueballs in QCD. Its configurations exhibit frequently
properties of magnetic monopoles. Electric charges, if they are not screened,
act on each other by forces that are independent on the distance, which means
that the confinement takes place.  The vacuum has a complex structure, as it
contains a condensate of the monopoles.  This explains the confinement as a
dual Meissner effect from superconductivity.

If charged fermions are present, the vacuum also contains a fermion
condensate. If the bare fermion mass is zero, the corrresponding chiral
symmetry is spontaneously broken, fermions acquire mass, and a massless
pseudoscalar Goldstone boson is present. It is an example of a dynamical
symmetry breaking, i.e. without help of a Higgs field.

The conclusion is, that compact abelian gauge theories on the lattice have (at
least) two phases with very different properties. For large $\beta$, it is the
Couplomb phase, where everything we have ever thought about abelian theories
holds. For small $\beta$ it is the phase with confinement and other properties
resembling in many aspects QCD. Both phases are separated by a phase
transition expected in the region in which both expansions cease to be
applicable, i.e. at
\begin{equation}
             \beta\; \simeq \; \beta^c \; = \; O(1).
  \label{betac}
\end{equation}
Its existence follows from the fact that it is not possible to connect
analytically phases which are so different.

The crucial question concerning the confinement phase is, whether its
properties are only an artefact of the lattice regularization or whether it is
possible to construct here a continuum limit preserving these properties.  The
second case would mean the existence of a continuum abelian gauge theory,
defined by this limit, which has not yet been formulated as a lagrangian
theory directly in continuum. From the point of view of quantum field theory
this is a very interesting possibility.

As follows from the considerations in the preceeding chapter, for such a
construction we need a critical point. The just described phase transition
between the confinement and Coulomb phases is an obvious candidate, and it is
thus necessary to assess its properties. However, this phase transition is
outside the region of reliability of analytic methods which are available. Its
position (\ref{betac}) is just ``between'' these regions and its study is
reserved for numerical methods. By means of these methods it is easy to verify
its existence and to determine its position (in agreement with the expectation
(\ref{betac})). But it is very difficult (for reasons well known to
statistical physicists) to decide whether it is a continuous phase transition
or a ``weak'' first order transition.  Results of numerical simulations are
not conclusive. Many our results suggest second order [1-6],
%\cite{JeLa96a,JeLa96b,CoFr97b,CoJe99,JeNe99,CoFr98b}, 
and some of them we shall see in a moment, but the question remains open.

In the cases of the pure gauge field or with a charged fermion field, no other
phase transition is available. This means that it is still not clear whether
the confinement phase exists in these theories in the continuum limit.
Unanswered remains also the question of the Landau pole in QED. Its study on
the lattice has to be performed in the Coulomb phase, where an approach to the
Landau pole corresponds to the approach to $\beta^c$. Here the perturbative
approach cesses to be valid. Should the transition be of second order, and if
in addition certain scaling behaviour holds, it would imply that actually QED
does not have a Landau pole. In my opinion, all attempts to clarify the
situation in numerical simulations do not yet allow a reliable statement about
the short distance properties of QED.

Do there exist abelian lattice gauge theories with some further phase
transitions?  One new possibility arises in the theory with a charged scalar
field. This is the scalar QED including a selfcoupling of the scalar field,
i.e. the U(1) Higgs model. This model has in a broad interval of $\beta$ also
the Higgs phase transition. Several research groups, including ours,
investigated whether it is not possible to construct also a continuum limit
different from that which agrees with the perturbation theory. If this were
so, it would substantially influence our understanding of the Higgs mechanism.
These nonperturbative studies confirmed quite reliably that the perturbative
understanding of the Higgs mechanism is sufficient. This means, in particular,
that there exists an upper energy bound for the validity of the elecroweak
sector of the Standard Model. This bound has been calculated both
perturbatively and numerically and decreases with the growing Higgs boson
mass.

Unexpected and interesting nonperturbative results have been obtained in the
abelian lattice gauge theory containing simultaneously fermion and scalar
fields of the same charge, called \chup$_4$ model [7-10].
%\cite{FrJe95a,FrFr95,FrJe98b,FrJe98a}. 
In the confinement phase with
dynamically broken chiral symmetry the scalar field does not condensate (thus
it is not a Higgs field in this phase!), but suppresses the chiral condensate,
the more the lighter it is. (For experts: imagine QCD with an additional
scalar ``quark''.) For certain values of the parameters of the theory chiral
symmetry gets restored through a phase transition.  For small values of
$\beta$ this new phase transition is doubtlessly of second order, and thus it
is possible to consider the continuum limit here.

We have studied this phase transition in the \chup$_4$ model with considerable
effort. One of the numerical results, expected on the basis of the strong
coupling expansion in powers of $\beta$, is that, for $\beta < 0.66$, this
transition belongs to the same universality class as the Nambu--Jona-Lasinio
(NJL) model of the four-fermion interaction.  Here the \chup$_4$ model is
practically equivalent to this well known model, accessible to analytical
studies by means of the Schwinger-Dyson equations. We have found, however,
that at $\beta \simeq 0.66$ (again so far away that the strong coupling
expansion is not applicable) there exists a particular type of phase
transition, the so-called tricritical point. Here it is possible to construct
a continuum limit which is essentially different from the NJL model. It is a
new type (universality class) of dynamical chiral symmetry breaking and mass
generation, not yet predicted by any analytic method. We shall describe it in
more detail later.

Since this model is interesting from the general point of view of quantum
field theory, we have investigated it also in other spacetime dimensions.  In
two dimensions \cite{FrJe96c} the \chup$_2$ model belongs to the same
universality class as the popular Gross-Neveu model in the same dimension.
This means, for example, that it is asymptotically free. In three dimensions
\cite{BaFo98} the \chup$_3$ model at small values of $\beta$ has a continuum
limit belonging to the universality class of the Gross-Neveu model in three
dimensions and is thus nonperturbatively renormalizable. For intermediate and
even large values of $\beta$ the properties of the \chup$_3$ model remain
rather mysterious. Perturbative expansion does not hold. Here the \chup$_3$
model merits further nonperturbative investigation, in particular because it
is similar to some models proposed for the high temperature superconductivity
\cite{Je01}.

In the next three chapters we shall illustrate some of the described phenomena
by several concrete numerical results. We shall skip all the technical
details, however, as these can be easily found in the cited papers. These
papers also contain references to many important works of other authors.

%%%%%%%%%%%%%%%%%%%%%%%%%%%%%%%%%%%%%%%%%%%%%%%%%%%%%%%%%%%%%%%%%%%%%%%%
\section{Pure U(1) gauge field on the lattice}
%%%%%%%%%%%%%%%%%%%%%%%%%%%%%%%%%%%%%%%%%%%%%%%%%%%%%%%%%%%%%%%%%%%%%%%%

In any quantum field theory with cutoff the choice of action is not unique
because it is possible to add rather arbitrarily terms proportional to the
inverse powers of this cutoff. Such a nonuniqueness allows us to choose the
action with some particular virtues. For example, in the case of the U(1)
lattice gauge field instead of (\ref{S}) one can use other periodic functions
of $\Theta_p$ with the expansion (\ref{Sexpanded}). The choice of the periodic
Gaussian function (Villain action) allows a very simple duality transformation
of the gauge field, which reveals the second nature of the U(1) gauge field
with compact link variables (\ref{U}): its nonperturbative properties are
governed by magnetic monopoles arising as a consequence of the selfinteraction
(\ref{Sexpanded}) as topologically nontrivial configurations of the gauge
field.

\begin{figure}[t]
\begin{center}
  \epsfig{file=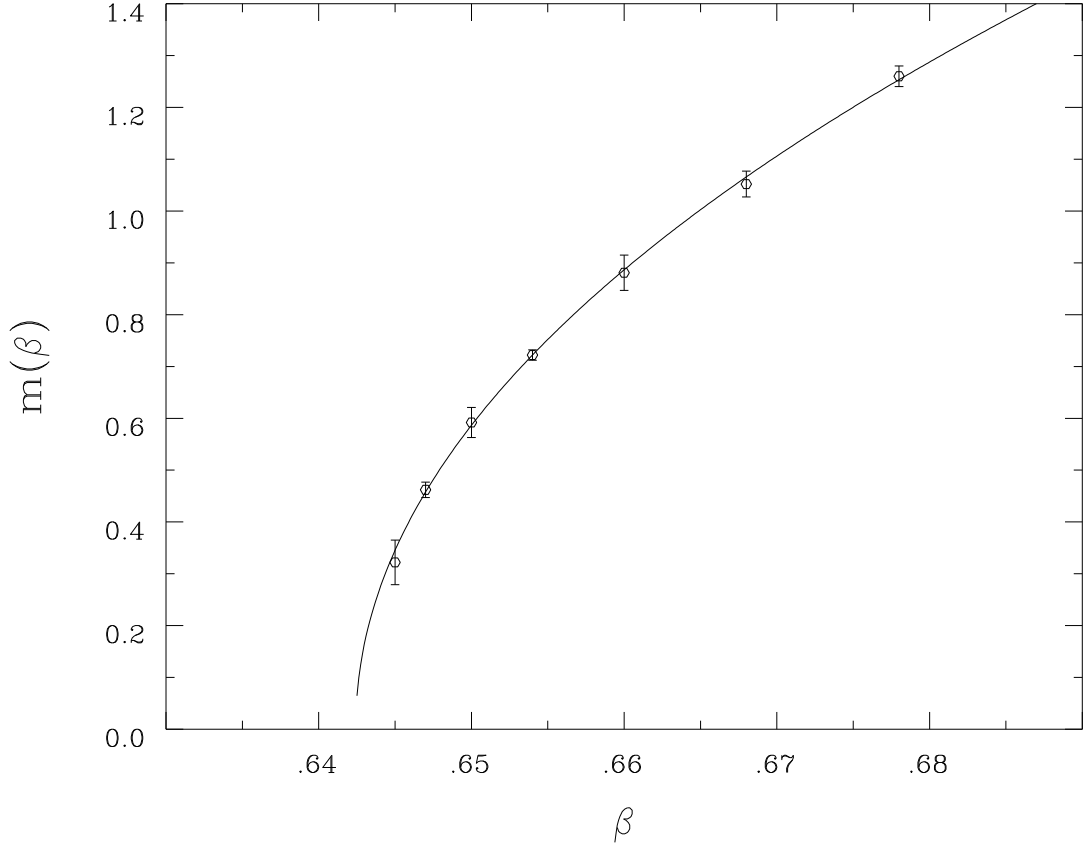,angle=90,width=\hsize}
  \caption{
    \label{monoscale}
    Scaling behaviour of the type (\ref{scalpot}) of the magnetic monopole
      mass in the vicinity of the phase transition in the Coulomb phase of the
      pure U(1) gauge theory. (Taken from Ref.~\cite{JeNe99}.)}
\end{center}
\end{figure}

\begin{figure}[t]
\begin{center}
  \epsfig{file=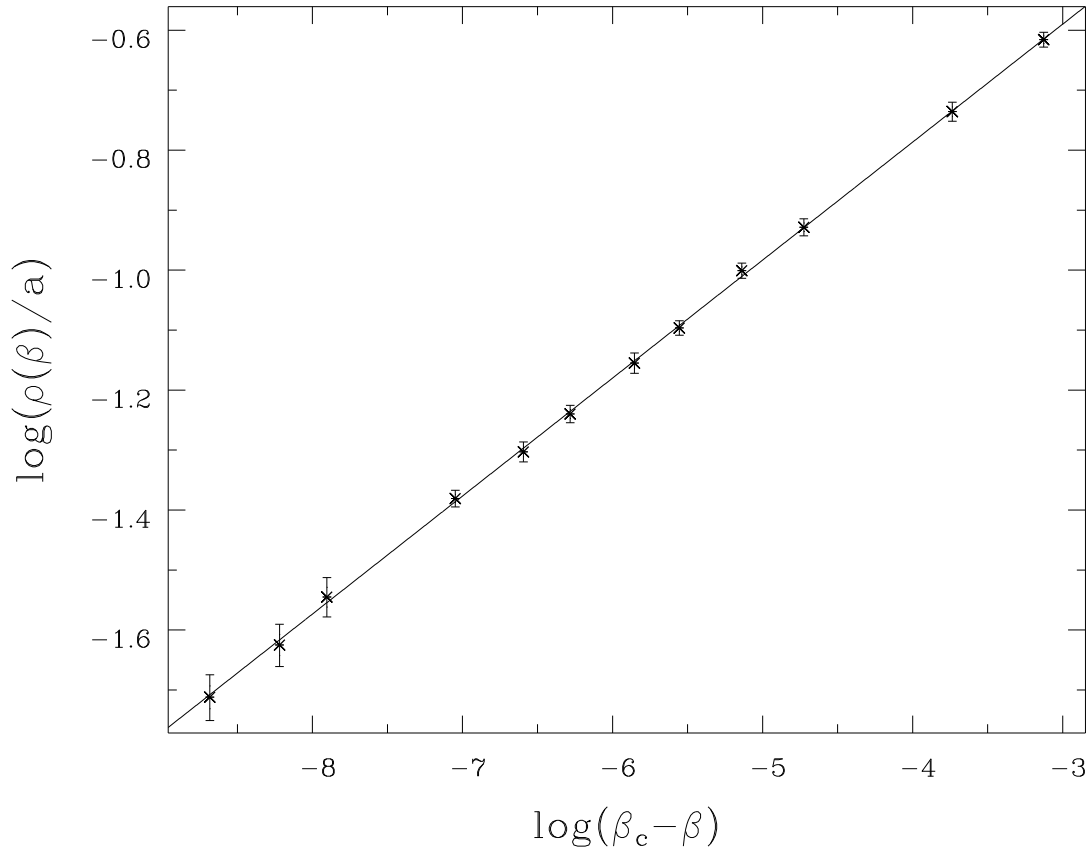,angle=90,width=\hsize}
  \caption{
    \label{condlog} 
    Scaling behaviour of the type (\ref{scalpot}) of the magnetic monopole
      condensate in the vicinity of the phase transition in the confinement
      phase of the pure U(1) gauge theory. The logarithmic scales on both axes
      stressthe remarkable accuracy of this behaviour. (Taken from
      Ref.~\cite{JeNe99}.) }
\end{center}
\end{figure}

Properties of these monopoles in the vicinity of the phase transition must be
studied numerically \cite{JeNe99}. Their most interesting property is the
scaling behaviour. In Figs.~\ref{monoscale} and \ref{condlog} we show some
numerical data and the fits using formula (\ref{scalpot}). Similar, though not
that accurate, is the scaling behaviour (\ref{scalpot}) of several gauge balls
in the confinement phase \cite{CoFr97b,CoJe99} and also of some thermodynamic
quantities \cite{JeLa96a,JeLa96b}.

%%%%%%%%%%%%%%%%%%%%%%%%%%%%%%%%%
%%%%%%%%%%%%%%%%%%%%%%%%%%%%%%%
%

The observed validity of the scaling behaviour is a support for the
conjecture, that in the case of the pure gauge theory it is possible to
construct a continuum limit, and thus an interesting nonperturbatively defined
renormalized U(1) gauge theory, in which the magnetic monopoles play an
important role.

The problem is, however, that this conjecture is based only on the numerical
data, which can be obtained only in a limited interval of the correlation
lengths $\xi_i$. It is not possible to prove numerically the validity of the
scaling behaviour in the limit (\ref{xi}). Nor is it possible to disprove
this hypothesis, since even if one succeeded to demonstrate, that for certain
choice of the action the phase transition is of first order, it would not
exclude that there exists another, more suitable choice, where the continuum
limit really exists.

These considerations illustrate the fundamental difficulty of the
nonperturbative studies on the lattice of field theories which are not
asymptotically free. Without reliable predictions about the existence and
properties of critical points they have only a suggestive character.

Situation analogous to the pure U(1) gauge theory arises also in the vicinity
of $\beta^c$ in the lattice U(1) gauge theory with fermions, i.e. compact
lattice QED \cite{CoFr98b}. The difference is that because of the
expensiveness of calculations with fermions the data are much less precise and
presently even a mere suggestion is not yet apparent. The problem of the
Landau pole in QED remains open. This is so even if in the noncompact
formulation of QED the existence of the Landau pole should be confirmed, as
the compact formulation might provide a more appropriate nonperturbative
definition of QED.

%%%%%%%%%%%%%%%%%%%%%%%%%%%%%%%%%%%%%%%%%%%%%%%%%%%%%%%%%%%%%%%%%%%%%%%%
\section{\chup$_4$ model}
%%%%%%%%%%%%%%%%%%%%%%%%%%%%%%%%%%%%%%%%%%%%%%%%%%%%%%%%%%%%%%%%%%%%%%%%

\begin{figure*}
  \begin{center}
    \epsfig{file=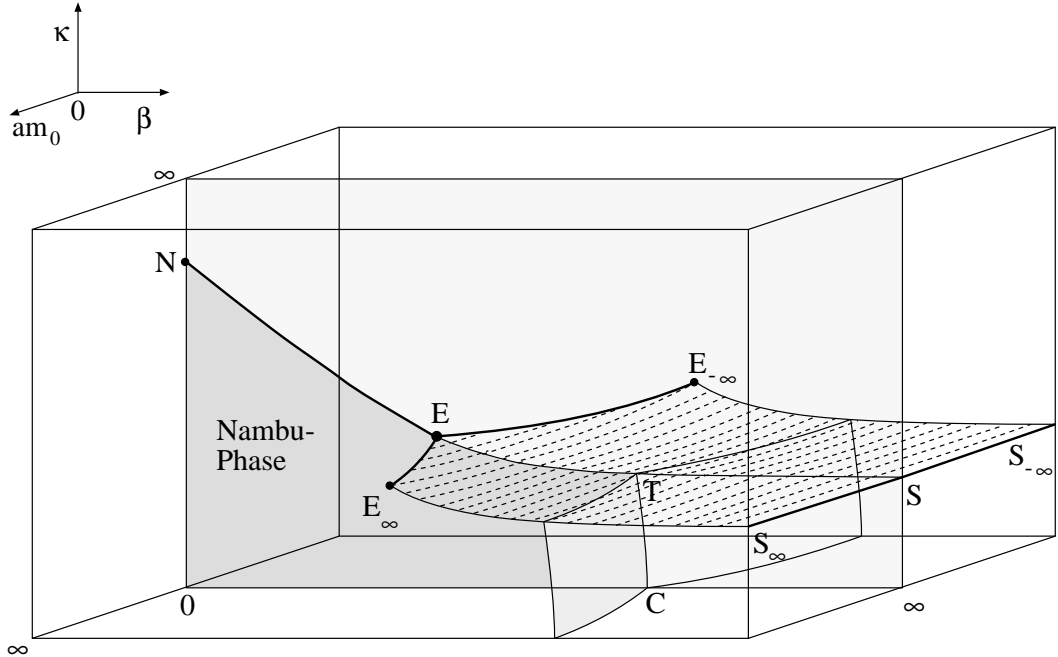,width=.85\hsize}%
    \caption[xxx]{%
      Schematic phase diagram of the \chup$_4$ model. Numerical data strongly
      suggest that the three critical lines NE, E$_\infty$E and E$_{-\infty}$E
      intersect in one, tricritical point E. (Taken from
      Ref.~\cite{FrJe98b}.)}
    \label{fig:pd4d3}
  \end{center}
\end{figure*}

This U(1) lattice gauge theory \cite{FrJe95a} contains simultaneously a
fermion field $\chi$ and a scalar field $\phi$ of the same charge.  A Yukawa
coupling between the fields $\chi$ and $\phi$ is absent due to the charge
conservation. Still, the model is quite complicated and contains several
parameters. The most important are $\beta$, given by Eq.~(\ref{beta}), bare
mass $am_0$ of the field $\chi$ in the units of $a$, and a dimensionless
parameter $\kappa \in \langle 0,\infty)$ which is a monotonous function of the
bare mass squared $m^2_{\phi,0}$ of the scalar field $\phi$ such that
\begin{equation}
\kappa = 0 \; \leftrightarrow m^2_{\phi,0} = +\infty, \;\;\;
 \kappa = \infty \; \leftrightarrow m^2_{\phi,0} = -\infty.
  \label{kappa}
\end{equation}
For $\kappa \simeq 0$ the scalar field is negligible, whereas for larger
values of $\kappa$ this field condenses and substantially influences the
properties of the model. The phase diagram, which we have investigated in some
detail [7-9],
%\cite{FrJe95a,FrFr95,FrJe98b}, 
is shown schematically in
Fig.~\ref{fig:pd4d3}. In the plane $am_0=0$ the model has a global
U(1)$_L\otimes$U(1)$_R$ chiral symmetry.

%FFFFFFFFFFFFFFFFFFF Fig. 3 FFFFFFFFFFFFFFFFFFFFFFFFFFFFFFFFFFFFFFFFFF

Many properties of the \chup$_4$ model can be deduced if we imagine the model
as a compact lattice QED with added field $\phi$, or as a U(1) Higgs model
with added field $\chi$.  In the first case it is clear that similar to QED
the model has, for small $\beta$ and $\kappa$ in the chiral limit $am_0=0$, a
phase of spontaneously broken chiral symmetry, which we call Nambu phase.  In
this phase the charges are confined. As follows from the relation
(\ref{kappa}), $m^2_{\phi,0}$ is large positive. With increasing $\kappa$ the
influence of the scalar field increases as $m^2_{\phi,0}$ decreases. Therefore
the chiral condensate decreases, and for certain values of $\kappa$ we observe
a second order phase transition, indicated in Fig.~\ref{fig:pd4d3} by the line
NE.

If we start with the U(1) Higgs model, we conclude that for any $am_0 \in
(-\infty, +\infty)$ and sufficiently large $\beta$ there exists a Higgs phase
transition of the first order, indicated in Fig.~\ref{fig:pd4d3} by the
approximately horizontal surface. It is known that the Higgs transition in the
Higgs model ends around $\beta = O(1)$ with a critical point. These points for
different $am_0$ form the critical lines E$_\infty$E and E$_{-\infty}$E
roughly in the $am_0$ direction.

By extensive numerical investigation we have found a solid evidence that these
three critical lines intersect at one point E. As follows from the theory of
such tricritical points in statistical physics, scaling behaviour in their
vicinity is substantially different from that in the vicinity of the critical
lines. This means that at the point E the model has a continuum limit which is
different from that obtained on the line NE, where the NJL model is obtained.

The existence and properties of the point E do not yet have any analytic
explanation. From the scaling behaviour indicated by the numerical studies the
continuum limit constructed at the point E from the Nambu phase might have the
following properties \cite{FrJe98b,FrJe98a}:
\begin{itemize}
\item Confinement of both charged fields $\chi$ and $\phi$, which means that
  the spectrum does not contain particles belonging to these fields.
\item Dynamical breaking of the chiral symmetry U(1)$_L\otimes$U(1)$_R$ and
  nonzero fermion condensate $\langle \overline \chi\chi \rangle$.
\item Goldstone boson $\pi$ (``pion'') of the mass $m_\pi$ and a coupling
  constant $f_\pi \neq 0$.
\item PCAC relation $(am_\pi)^2 \propto am_0$.
\item Neutral fermion $F$ in the spectrum with a nonzero mass $m_F$, whose
  field is composed of both charged fields,
   \begin{equation}
               F(x) \; = \; \phi^+(x) \chi(x).
     \label{F}
   \end{equation}
 \item Scalar gaugeball $S$ of mass $m_S
   \simeq m_F/2$ in the spectrum.
\item Effective Yukawa coupling $F\pi\pi$, whose strength achieves the upper
  bound of this interaction.
\item Vector boson (``$\rho$'') composed of $\overline \chi$ and $\chi$. But
  the analogous scalar boson (``$\sigma$'') has not been observed.
\end{itemize}

If we imagine, that for example the subgroup U(1)$_L$ of the broken chiral
symmetry is gauged, then the corresponding U(1)$_L$ chiral gauge symmetry is
broken by the fermion condensate $\langle \overline \chi\chi \rangle$, and the
vector boson corresponding to this gauge field is massive. This example
together with the listed properties show that the \chup$_4$ model has
properties similar to the Higgs-Yukawa sector in the spontaneously broken
chiral gauge theories with fermions (like the Standard Model).

The \chup$_4$ model in the vicinity of the point E is thus a prototype of a
possible nonperturbative alternative to the Higgs-Yukawa mechanism of the mass
generation.  Its disadvantage is a great dynamical complexity: the new
vectorlike U(1) gauge theory introduced for this purpose would have to be
strongly interacting in order to induce the required dynamical breaking of the
chiral symmetry.

Such modifications of the Standard Model have been considered for example in
connection with the idea of top-quark condensate caused by a new strong gauge
interaction. A weird property from the point of view of possible experimental
distinction between both mechanisms is the fact that though our model does not
have a Higgs boson (``$\sigma$''), it has a new scalar $S$. However, this
scalar is only weakly coupled to fermions, hopefully being thus
distinguishable from the Higgs boson.

%%%%%%%%%%%%%%%%%%%%%%%%%%%%%%%%%%%%%%%%%%%%%%%%%%%%%%%%%%%%%%%%%%%%%%%%
\section{\chup$_3$ model}
%%%%%%%%%%%%%%%%%%%%%%%%%%%%%%%%%%%%%%%%%%%%%%%%%%%%%%%%%%%%%%%%%%%%%%%%

A model analogous to that described in the previous section, but now in three
dimensions, is the \chup$_3$ model \cite{BaFo98}. For small $\beta$ its
scaling behaviour is very similar to that of the Gross-Neveu model in three
dimensions, and it is thus presumably nonperturbatively renormalizable in the
same sense as that model. Thus it possesses a continuum limit with two phases,
one with broken symmetry, which is a certain analogue to the chiral symmetry,
(Nambu phase), and one phase with this symmetry restored (Higgs phase). Though
these results are remarkable for a strongly coupled gauge theory, they also
mean that the \chup$_3$ model for small $\beta$ is nothing really new.

\begin{figure*}
  \begin{center}
%    \leavevmode  
    \epsfig{file=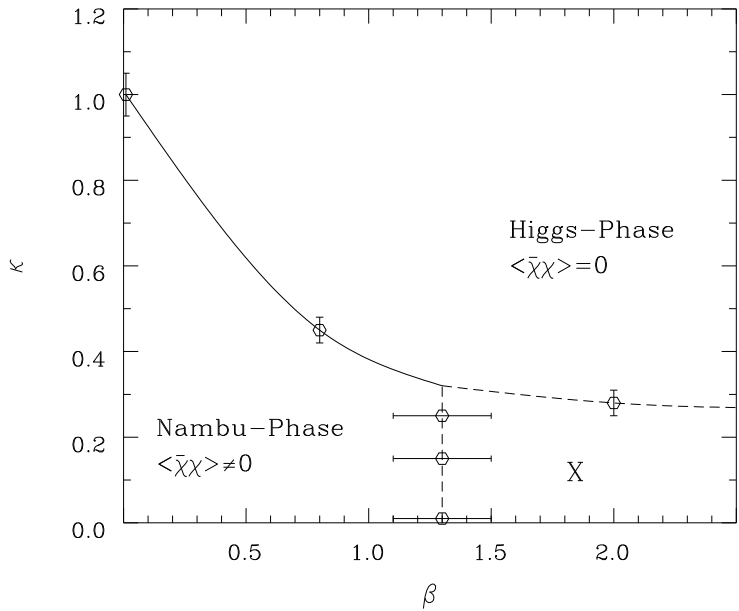,angle=90,width=0.85\hsize}%\hspace*{-2cm}%
    \vspace{-10mm}%
    \caption[xxx]{%
      Phase diagram of the \chup$_3$ model at $m_0=0$. For $\beta < 1$ the
      chiral phase transition between the Nambu and Higgs pahases is
      clearly seen. The continuation of this transition at $\beta > 1$ is
      unclear, however, and threrefore we dop not know whether in the region X
      teh symmetry is broken or not.  (Taken from Ref.~\cite{BaFo98}.)}
    \label{fig:pd}
  \end{center}
\end{figure*}

For intermediate and even large $\beta$ the situation is different. The
current status of our understanding of the \chup$_3$ model in the chiral
limit $am_0=0$ is described by Fig.~\ref{fig:pd}. We show two possibilities
for the position of the chiral phase transition for $\beta>1$, as we currently
cannot decide which one is correct. Therefore we do not know whether in the
region denoted X the symmetry is broken or not.

The substance of the problem is very interesting.  Because it concerns a
region with small $\kappa$, one can neglect the scalar field nearly in the
whole region X. Then the question is what are the properties of compact
QED$_3$ at large $\beta$ and one would expect that these are well known. This
is not the case, however, because the perturbation expansion fails to grasp
important properties of that theory even for $\beta \rightarrow \infty$.
Already in the case of pure gauge field one finds even for arbitrarily large
$\beta$ confinement of static charges by a constant force.

On the basis of analytic arguments it is expected that if $N_F$ charged
fermions are added, then for small $N_F < N_F^c$ the chiral symmetry is broken
in the whole range of $\beta$ including $\beta \rightarrow \infty$, whereas
for $N_F > N_F^c$ the symmetry gets restored with increasing $\beta$ at some
finite $\beta$. This are properties similar to QCD$_4$. For QED$_3$ one
expects $ N_F^c \simeq 3-4$.

However, in calculations with $ N_F = 2$ we have found for $\beta \simeq 1.3$
an indication of a phase transition (vertical line in Fig.~\ref{fig:pd}),
which would mean a substantially smaller value of $N_F^c$ than expected.
Naturally, because of the small sizes of our lattices, we cannot exclude that
the condensate rapidly but analytically decreases around $\beta \simeq 1.3$ to
a small but nonvanishing value. It is very difficult to distinguish
numerically such a decrease from a genuine phase transition. Therefore the
chiral phase transition could also take place on the horizontal line in
Fig.~\ref{fig:pd}.  Apart from these possibilities it seems also plausible
that, as in the \chup$_4$ model, there is a tricritical point around $\beta
\simeq 1$. We have not yet started a search for it.

The questions of the position of the chiral phase transition and existence of
a tricritical point in the \chup$_3$ model on the lattice are not merely
technical, as they decide about the existence of the continuum limit and about
properties of the corresponding abelian gauge theory in three dimensions,
about the existence of fixed points in such a theory, etc. Apart from the
importance for the general quantum field theory, these properties might be
relevant in connection with the high temperature superconductivity
\cite{Je01}. There are models for this phenomenon based on strongly coupled
gauge theory with analogies to the fields $\chi$ and $\phi$, called spinons
and holons. A better understanding of the \chup$_3$ model on the lattice might
contribute to a dynamical justification or rejection of such models.

%%%%%%%%%%%%%%%%%%%%%%%%%%%%%%%%%%%%%%%%%%%%%%%%%%%%%%%%%%%%%%%%%%%%%%%%
\section{Conclusion}
%%%%%%%%%%%%%%%%%%%%%%%%%%%%%%%%%%%%%%%%%%%%%%%%%%%%%%%%%%%%%%%%%%%%%%%%

Numerical studies of abelian lattice gauge theories indicates that new
renormalized abelian gauge theories might exist, with properties quite
different from abelian theories defined by means of the usual perturbative
expansion.  Because of the coupling strength O(1) in such theories,
confinement and dynamical breaking of both global and local symmetries at
short distances would appear. A dynamical generation of particle masses, both
of bosons and fermions would take place. Such theories might, in principle,
constitute an alternative to the Higgs-Yukawa mechanism.

Of course, the fundamental difficulty of the numerical investigation is the
inherent restriction of the results on the nonvanishing values of the lattice
constant $a$. The observed scaling behaviour has to be extrapolated to the
limit $a \rightarrow 0$ without a theoretical justification of such an
extrapolation.  The described results thus are not conclusive but only
suggestive.  Nevertheless, they can, as it is sometimes useful in quantum field
theory, inspire new hypotheses about properties of gauge field theories which
are not asymptotically free.  In any case, the observed scaling behaviour can
be used for a construction of effective field theories valid in a finite
interval of the values of a cutoff. That this is not a small achievement is
illustrated by the highly priced current theory of electroweak interactions,
which is ``merely'' an effective theory, too.

Could such speculations about nonperturbative abelian gauge theories be
relevant phenomenologically? Probably not if the Higgs boson and
supersymmetric particles will be confirmed experimentally, because then the
further development will continue on the perturbative road according to the
existing plans. However, plans sometimes do not get fulfilled.

\vspace{1cm}
{\bf Acknowledgement.}

This lecture has been delivered in Prague in honor of the memory of Prof. V.
Votruba. I thank J. Ho\v{r}ej\v{s}\'{\i} and other organizers of this meeting
for warm hospitality. I am grateful to J. Ho\v{s}ek for discussions on the
dynamical symmetry breaking and for drawing my attention to a relationship
between our work and some models of high temperature superconductivity.

\end{document}